\documentclass[onecolumn,showpacs,nobibnotes,nofootinbib,12pt]{revtex4-1}
\usepackage{amsmath,amssymb}
\usepackage{wasysym}
\usepackage{graphicx}
\usepackage{color}

\usepackage{epstopdf}
\usepackage[pdftex]{hyperref}

\newcommand{\td}{\text{d}}

\begin{document}
\author{Yan-Qing Ma}
\email{yqma@pku.edu.cn}
\affiliation{School of Physics and State Key Laboratory of Nuclear Physics and Technology, Peking University, Beijing 100871, China}
\affiliation{Center for High Energy Physics, Peking University, Beijing 100871, China}
\affiliation{Collaborative Innovation Center of Quantum Matter, Beijing 100871, China}

\author{Jian-Wei Qiu}
\email{jqiu@jlab.org}
\affiliation{Theory Center, Jefferson Lab, Newport News, VA 23606, USA}

\author{Hong Zhang}
\email{zhang.5676@osu.edu}
\affiliation{Department of Physics, The Ohio State University, Columbus, Ohio 43210, USA}

\title{Rotation-invariant observables in polarization measurements}

\begin{abstract}
Polarization measurements provide a detailed method to test the Standard Model and to search for new physics.
Most previous studies depend on pre-selected coordinates, which blurs the significance of the results. The construction of two rotation-invariant observables in vector boson decay into a fermion pair has been proved to be a big success. In this work, we show that there are more rotation-invariant observables and provide a general recipe to find all of them in an arbitrary decay process. Taking spin-1/2 and spin-1 particle decay processes as examples, we calculate the explicit expressions of all rotation-invariant observables, which can serve as a robust test of the detector acceptance and help the analysis of experimental data.
\end{abstract}
\maketitle

\allowdisplaybreaks


\section{Introduction}

Studying polarization of particles produced in high-energy collisions provides  more information of a certain process. It can serve as a powerful tool to test the Standard Model, as well as to search for new physics. All polarization measurements depend on a pre-selected frame. Consequently, the outputs are inevitably dependent on the frame choice. Such dependence often causes cumbersomeness in comparison between theoretical predictions and experimental measurements, as well as comparison between different measurements. An example is the study of $J/\psi$ polarization, in which the results from the Tevatron and the LHC seem to be inconsistent \cite{Braaten:2014ata}.

Recently, a few rotation-invariant observables have been proposed \cite{Faccioli:2010ej, Faccioli:2010ps, Palestini:2010xu, Shao:2012fs}, based on the fact that all experimentally interesting frames are related by a rotation in the production plane \cite{Braaten:2014ata}. These rotation-invariant observables provide much more powerful test for the underline production mechanism, and they also provide a non-trivial check of the unaddressed systematic uncertainties for experimental data analyses \cite{Faccioli:2010kd,Chatrchyan:2012woa,Chatrchyan:2013cla}.

In this work, we show that there could be more rotation-invariant observables and provide a general recipe to find all of them for an arbitrary decay process. The rest of this paper is organized as follows. In section~\ref{sec:general}, we show that the angular distribution of the decay products from a spin-$J$ particle can be expanded by spherical harmonics $Y_{l,m}$ with $l\leq 2J$. Then in section~\ref{sec:RI}, we introduce a general method of finding all the rotation-invariant observables. In section~\ref{sec:application}, we apply our method to obtain explicit expressions of the rotation-invariant quantities for three most phenomenologically important cases, i.e. the decay process of a particle with spin $1/2$, $1$ and $2$, respectively.


\section{General Analyses}\label{sec:general}

\begin{figure}
\begin{center}
\includegraphics[width=0.21\textwidth]{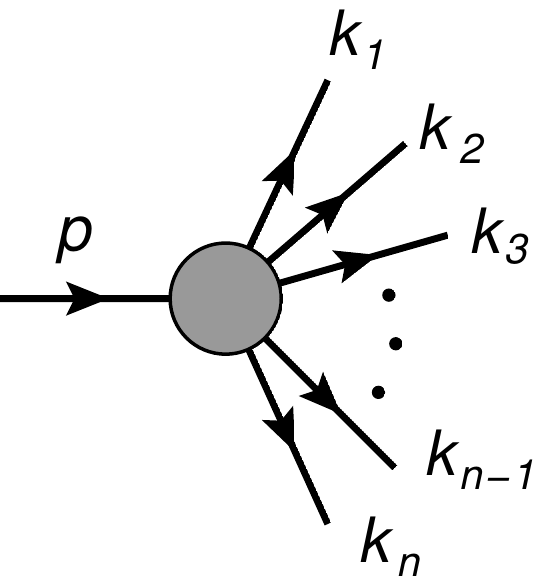}
\end{center}
\caption{A general decay process from a parent particle with momentum $p$ to $n$ daughter particles. }
\label{fig:decay}
\end{figure}

Let us first consider a vector boson $V$ with mass $M_V$ decaying into $n$ particles, as shown in Fig.~\ref{fig:decay}.  Angular distribution of a daughter particle with momentum $k_1$ in the rest frame of the parent particle $V$ can be expressed as
\begin{align}\label{eq:Vdistribution}
\begin{split}
\frac{\td\Gamma}{\td \Omega_1}=&\frac{1}{2M_V}
\Big\{\int \frac{|\vec{k}_1|^2 \td |\vec{k}_1|}{(2\pi)^3 2E_1}~ \prod_{l=2}^n\int \frac{\td^3k_l}{(2\pi)^3 2E_l} \Big[ \overline{|\mathcal{M}|^2}\Big]_{ij}(2\pi)^4\delta^{(4)}(p-\sum_{l=1}^n k_l)\Big\}\\
&\times \sum_{\lambda,\lambda'}\rho_{\lambda \lambda'}\epsilon^{i}_{\lambda}\epsilon^{*j}_{\lambda'},
\end{split}
\end{align}
where $\overline{|\mathcal{M}|^2}$ is the squared amplitude of the decay process with summation over spins of all decay products, $p^\mu=(M_V,0,0,0)$ is the momentum of $V$ in its rest frame, $\rho$ is the spin density matrix of $V$, and $\epsilon^{i}_{\lambda}$ ($\epsilon^{*j}_{\lambda'}$) are polarization vectors with polarization $\lambda$ ($\lambda'$) for $V$ in the amplitude (the complex conjugate of the amplitude).

After the integration over $\vec{k}_i$~(i$\geq$2) and $|\vec{k}_1|$, the only vector left in the curly bracket in Eq.~\eqref{eq:Vdistribution} is $\vec{n}_1=(n_x,n_y,n_z)=(\sin \theta\cos\phi, \sin\theta \sin \phi, \cos \theta)$, where $\theta$ and $\phi$ are the polar and azimuthal angles of $\vec{k}_1$ in the rest frame of $V$, respectively for a given choice of the coordinate system. As a result, the only possible tensor structures in the curly bracket are $\delta^{ij}$, $n_1^i n_1^j$ and $\varepsilon_{ijr}n_1^r$, where $i,j,r=1,2,3$. Therefore the left-hand side (LHS) of Eq.~\eqref{eq:Vdistribution} can be expanded by spherical harmonics $Y_{lm}(\theta,\phi)$ with $l\leq 2$. Especially, if the decay process conserves parity, terms proportional to $\varepsilon_{ijr}n_1^r$ are equal to zero, and then, the LHS of Eq.~\eqref{eq:Vdistribution} can be expanded by $Y_{lm}(\theta,\phi)$ with $l=0, 2$ only.

The same conclusion can be obtained from the transformation of Eq.~\eqref{eq:Vdistribution} under $SO(3)$ rotation. If we rotate the reference frame (passive interpretation), both polarization vectors on the right-hand side (RHS) of Eq.~\eqref{eq:Vdistribution} transform as the $J=1$ representation $D^{(1)}$ of $SO(3)$. Then the curly bracket on the RHS of Eq.~\eqref{eq:Vdistribution} must transform as $D^{(1)}\otimes D^{(1)}=D^{(0)}\oplus D^{(1)}\oplus D^{(2)}$. Therefore it can be expressed by linear combination of $Y_{lm}(\theta,\phi)$ with $l\leq 2$.

The argument with the rotational symmetry is very general to be applied to parent particles with any spin. With some algebra, one can show that the angular distribution of any daughter particle in the rest frame of the parent particle can always be expressed as
\begin{equation}\label{eq:YExpand}
\frac{1}{\Gamma}\frac{\td \Gamma}{\td \Omega}\equiv f(\theta,\phi)=\sum_{l=0}^{2J}\sum_{m=-l}^l f_{l,m}Y_{lm}(\theta,\phi),
\end{equation}
where $J$ is the spin of the parent particle. Since ${\td \Gamma}/{\td \Omega}$ is real, there are relations $f_{l,-m}=(-1)^m f_{l,m}^*$ for any $l$ and $m$. Together with the trivial relation $f_{0,0}=\frac{1}{\sqrt{4\pi}}$ fixed by the normalization condition, the number of the degrees of freedom of all coefficients $f_{\,l,m}$ is $4J(J+1)$.


\section{Rotation-Invariant Observables}\label{sec:RI}

If we rotate the reference frame from the original one to a new one, $f_{l,m}$ defined in Eq.~\eqref{eq:YExpand} changes accordingly. Since $SO(3)$ rotation has 3 degrees of freedom (usually chosen as the Euler angles), one expects $4J(J+1)-3$ independent combinations of $f_{l,m}$ to be invariant under the $SO(3)$ rotation.\footnote{The only exception is for $J=\frac{1}{2}$ parent particle, which will be explained in Sec~\ref{subsec:1/2}.} From Eq.~\eqref{eq:YExpand}, $2J$ of these rotation-invariant combinations can be easily identified,
\begin{equation}\label{eq:RISO32}
U_l=\sum_{m=-l}^l |f_{l,m}|^2,~~~~~~~~l=1,~2,~\cdots,~2J,
\end{equation}
which are quadratic in $f_{l,m}$. Eq.~\eqref{eq:RISO32} is from the singlet representation of $D^{(l)}\otimes D^{(l)}$. More rotation-invariant combinations can be constructed with higher powers of $f_{l,m}$. For example, the singlet representation in $D^{(l)}\otimes D^{(l)}\otimes D^{(l)}$ gives a rotation-invariant observable cubic in $f_{l,m}$. A more direct way to obtain all of these high-power rotation-invariant observables is to calculate
\begin{equation}\label{eq:RISO3}
W_{n}=\int d\Omega \left[f(\theta,\phi) -\frac{1}{4\pi}\right]^n,~~~~~~~~n=2~,3~,\cdots.
\end{equation}
In this way, we can find a complete set of $SO(3)$ rotation-invariant observables.

In practice, the most commonly-used frames (such as s-channel helicity frame, Collins-Soper frame \cite{Collins:1977iv}, and Gottfried-Jackson frame \cite{Gottfried:1964nx}) can be related by a $SO(2)$ rotation in the production plane, which is usually chosen as the $x-z$ plane of the reference frame \cite{Faccioli:2010kd}. Since $SO(2)$ rotation has only one degree of freedom, one thus expects two additional  rotation-invariant observables.

To obtain the rotation-invariant observables under $SO(2)$ rotation in $x-z$ plane, it is better to express Eq.~\eqref{eq:YExpand} in bases of $\bar{Y}_{lm}(\theta,\phi)$, which is the eigenstates of $\hat{J}_y$,
\begin{equation}\label{eq:YExpand2}
\frac{1}{\Gamma}\frac{\td \Gamma}{\td \Omega_B}=\sum_{l=0}^{2J}\sum_{m=-l}^l g_{\,l,m}\bar{Y}_{lm}(\theta,\phi),
\end{equation}
where $g_{l,m}$ is related to $f_{l,m}$ defined in Eq.~\eqref{eq:YExpand} by
\begin{align}
g_{\,l,m}=&\sum_{m'=-l}^{l}f_{l,m'}~e^{-i\frac{\pi}{2}(m'-m)}d^{\,l}_{m,m'}(-\frac{\pi}{2}),
\end{align}
and the Wigner $d$-function is given by
\begin{align}
d^{\,l}_{m,m'}(-\frac{\pi}{2})= \sum_{\nu=\text{max}\{0,m-m'\}}^{\text{min}\{l+m,l-m'\}}\frac{(-1)^\nu\sqrt{(l+m)!(l-m)!(l+m')!(l-m')!}}{2^l(l+m-\nu)!(l-m'-\nu)!(\nu-m+m')!\nu!}\,.
\end{align}

If another frame is related to the current frame by a rotation of angle $\delta$ in the $x-z$ plane, the coefficients $g'_{l,m}$ in the expansion similar to Eq.~\eqref{eq:YExpand2} in the new frame is $g'_{l,m}=e^{im\delta}g_{l,m}$. Thus the following observables are invariant under $SO(2)$ rotation in the $x-z$ plane:
\begin{subequations}\label{eq:RISO2a}
\begin{align}
\begin{split}
T_{\,l,0}&=g_{l,0},
\end{split}\\
\begin{split}
T_{\,l,m}&=\left|g_{\,l,m}\right|^2,~~~~m=1,2,\cdots,l.
\end{split}
\end{align}
\end{subequations}
Eq.~\eqref{eq:RISO2a} gives $2J+1$ $SO(2)$ rotation-invariant observables, and one can construct even more of them by multiplying $g_{l,m}$ with different values of $m$. However, only two of them is independent of the $SO(3)$ rotation-invariant observables defined in Eqs.~\eqref{eq:RISO32} and \eqref{eq:RISO3}. We find Eq.~\eqref{eq:RISO2a} is adequate to give the two independent $SO(2)$ rotation invariants.


\section{Application}\label{sec:application}

In this section, we apply Eqs.~\eqref{eq:RISO32}, \eqref{eq:RISO3} and \eqref{eq:RISO2a} to the three most important cases in phenomenology, i.e. the decay of a particle with spin $1/2$, $1$ and $2$, respectively. We find explicit expressions for the $SO(3)$ and $SO(2)$ rotation-invariant observables. Some of these invariants have been found in previous literatures. We show that our formula can reproduce all of them, and also give more invariants which have not been realized before.  These explicit expressions could be used as a robust test of the unaccounted for systematic uncertainty in experimental measurements.

\subsection{Spin-${1}/{2}$ particle decay}\label{subsec:1/2}

The angular distribution of a daughter particle in the rest frame of the parent particle with spin-$1/2$ can be expressed as
\begin{align}\label{eq:jpsi}
\frac{1}{\Gamma}\frac{d\Gamma}{d \Omega}=&\frac{1}{4\pi} \Big[ 1+ 2A_\theta \cos \theta +2A_\phi \sin \theta \cos \phi + 2A_{\perp\phi} \sin\theta \sin\phi
 \Big].
\end{align}
Although there are only three degrees of freedom in this expression, from Eq.~\eqref{eq:RISO32} and Eq.~\eqref{eq:RISO3}, we can still obtain a $SO(3)$ rotation-invariant observable
\begin{align}\label{eq:halfSO3}
\begin{split}
U_1=W_2=
\frac{A_\theta^2+A_\phi^2+A_{\perp\phi}^2}{3\pi} .
\end{split}
\end{align}
As discussed in Sec.~\ref{sec:RI}, we expect $4J(J+1)-3$ independent combinations of $f_{l,m}$ to be invariant under the $SO(3)$ rotation.  This would lead to zero ($[4J(J+1)-3]_{J=1/2}=0$) independent combination of $f_{l,m}$ from the decay of a spin 1/2 particle to be invariant under the $SO(3)$ rotation.  Having the finite $U_1$ or $W_2$ in Eq.~(\ref{eq:halfSO3}) is not completely inconsistent because of the fact that it can be expressed as a linear combination of two $SO(2)$ rotation-invariants, $T_{1,0}^2$ and $T_{1,1}$, given below.

From Eq.~\eqref{eq:RISO2a} we obtain two $SO(2)$ rotation-invariant observables
\begin{align}\label{eq:halfSO2}
\begin{split}
T_{1,0}=\frac{A_{\perp\phi}}{\sqrt{3\pi}} \, ,
\end{split}
\end{align}
and
\begin{align}\label{eq:fullSO2}
\begin{split}
T_{1,1}=
\frac{1}{12\pi}\left[
A_\theta^2 + A_\phi^2 \right] .
\end{split}
\end{align}
In practice, the overall constant factors of these invariants could be dropped for convenience.

\subsection{Spin-$1$ particle decay}

The angular distribution of any daughter particle in the rest frame of a spin-1 parent particle is usually expressed as
\begin{align}\label{eq:jpsi}
\frac{1}{\Gamma}\frac{d\Gamma}{d \Omega}=&\frac{1}{4\pi(1+\frac{\lambda_\theta}{3})} \Big[ 1+ \lambda_\theta \cos^2\theta + \lambda_\phi \sin^2\theta \cos 2\phi + \lambda_{\theta\phi} \sin 2\theta \cos \phi + \lambda_{\perp\phi} \sin^2\theta \sin 2\phi
\nonumber\\
& \hspace{2cm}
+ \lambda_{\perp\theta\phi} \sin 2\theta \sin \phi
+2 A_\theta \cos \theta
+2A_\phi \sin \theta \cos \phi
+2A_{\perp\phi} \sin\theta \sin\phi
\Big].
\end{align}
The coefficients $A$'s equal to zero for the parity-conserving process. From Eq.~\eqref{eq:RISO32} and Eq.~\eqref{eq:RISO3}, we obtain $SO(3)$ rotation-invariant observables
\begin{subequations}\label{eq:jpsiRI3}
\begin{align}
\begin{split}
U_1&=\frac{3}{\pi}
\frac{A_\theta^2+A_\phi^2+A_{\perp\phi}^2}{(3+\lambda_\theta)^2},
\end{split}\\
\begin{split}
U_2&=\frac{1}{5\pi}
\frac{\lambda_\theta^2+3(\lambda_\phi^2+\lambda_{\theta\phi}^2+\lambda_{\perp\phi}^2+\lambda_{\perp\theta\phi}^2)}{(3+\lambda_\theta)^2},
\end{split}\\
\begin{split}
W_2&=U_1+U_2,
\end{split}\\
\begin{split}
W_3&=\frac{1}{70\pi^2(3+\lambda_\theta)^3}\Big[
(\lambda_\theta+3\lambda_\phi)(2\lambda_\theta^2-6\lambda_\theta \lambda_\phi +9\lambda_{\theta \phi}^2)\\
&\hspace{3cm}
+9(\lambda_\theta \lambda_{\perp\theta\phi}^2 -2 \lambda_\theta \lambda_{\perp\phi}^2 +6\lambda_{\theta\phi}\lambda_{\perp\theta\phi} \lambda_{\perp\phi} -3 \lambda_\phi \lambda_{\perp\theta\phi}^{2})\\
&\hspace{3cm}
+63\lambda_\theta(2A_\theta^2-A_\phi^2-{A_{\perp\phi}}^2)
+189 \lambda_\phi (A_\phi^2-A_{\perp\phi}^2)\\
&\hspace{3cm}
+378 (A_\theta A_\phi \lambda_{\theta\phi}+A_\theta A_{\perp\phi}\lambda_{\perp\theta\phi}+A_\phi A_{\perp\phi}\lambda_{\perp\phi})
\Big],
\end{split}\\
\begin{split}
W_4&=\frac{9}{20\pi}U_1^2+\frac{15}{28\pi}U_2^2+\frac{27}{14\pi}U_1 U_2\\
&\hspace{0.5cm}
+\frac{9}{35 \pi^3 (3+\lambda_\theta)^4}
\Big[A_\theta \lambda_\theta(A_\theta \lambda_\theta+6A_\phi \lambda_{\theta\phi}+6A_{\perp\phi}\lambda_{\perp\theta\phi}) - 12 A_\phi A_{\perp\phi}\lambda_\theta\lambda_{\perp\phi}\\
&\hspace{0.5cm}
+18A_\theta(A_\phi  \lambda_\phi\lambda_{\theta\phi}+A_\phi \lambda_{\perp\phi}\lambda_{\perp\theta\phi}+A_{\perp\phi}\lambda_{\perp\phi}\lambda_{\theta\phi}-A_{\perp\phi}\lambda_\phi\lambda_{\perp\theta\phi})
-9A_\theta^2(\lambda_\phi^2+\lambda_{\perp\phi}^2)\\
&\hspace{0.5cm}
-2\lambda_\theta^2(A_\phi^2+A_{\perp\phi}^2)
-6\lambda_\theta \lambda_\phi(A_\phi^2-A_{\perp\phi}^2)
-9(A_\phi \lambda_{\perp\theta\phi}-A_{\perp\phi}\lambda_{\theta\phi})^2
\Big],
\end{split}\\
\begin{split}
W_5&=\frac{5}{2\pi}\left(\frac{3}{7}U_1+\frac{5}{11}U2\right) W_3\\
&\hspace{0.5cm}
+\frac{3}{539 \pi^4 (3+\lambda_\theta)^5}
\Big\{
(A_\theta^2+A_\phi^2+A_{\perp\phi}^2)
\Big[\lambda_\theta^3+36 \lambda_\theta(\lambda_{\theta\phi}^2+\lambda_{\perp\theta\phi}^2)-261\lambda_\theta(\lambda_\phi^2+\lambda_{\perp\phi}^2)\\
&\hspace{0.5cm}
+297(\lambda_{\phi}\lambda_{\theta\phi}^2-\lambda_\phi \lambda_{\perp \theta\phi}^2+2\lambda_{\perp\phi}\lambda_{\theta\phi}\lambda_{\perp\theta\phi})
\Big]
+63\Big[\lambda_\theta^2+3(\lambda_{\phi}^2+\lambda_{\theta\phi}^2+\lambda_{\perp\phi}^2+\lambda_{\perp\theta\phi}^2)\Big]\\
&\hspace{0.5cm}
\times\Big(A_\theta^2\lambda_\theta+A_\phi^2\lambda_\phi-A_{\perp\phi}^2\lambda_\phi
+2A_\theta A_\phi \lambda_{\theta\phi}+2A_\theta A_{\perp\phi}\lambda_{\perp \theta\phi}+2A_\phi A_{\perp\phi}\lambda_{\perp\phi}\Big)
\Big\}.
\end{split}
\end{align}
\end{subequations}
Since there are eight real coefficients ($\lambda$'s and $A$'s) in Eq.~\eqref{eq:jpsi}, the quantities $U_{1,2}$ and $W_{3,4,5}$ are the only five independent $SO(3)$ rotation-invariant observables that we can construct. Any combination of them are also rotation invariant. From Eq.~\eqref{eq:RISO2a} we obtain the two $SO(2)$ rotation-invariant observables in $x-z$ plane,
\begin{subequations}\label{eq:jpsiRI2}
\begin{align}
\begin{split}
T_{2,0}=
-\frac{1}{2\sqrt{5\pi}}
\frac{\lambda_\theta+ 3\lambda_\phi }{3+\lambda_\theta},
\end{split}\\
\begin{split}
T_{2,2}=
\frac{3}{40\pi}\,\frac{(\lambda_\theta-\lambda_\phi)^2 + 4 \lambda_{\theta\phi}^2}{(3+\lambda_\theta)^2}.
\end{split}
\end{align}
\end{subequations}
In Eq.~\eqref{eq:jpsiRI2}, $T_{2,0}$ and $T_{2,2}$ are equivalent to the rotation-invariant observables obtained in Refs.~\cite{Faccioli:2010ps} and \cite{Palestini:2010xu}, respectively. $T_{2,0}$ is also equivalent to rotation-invariant observable $F_1$ defined in both Eq. (25) and Eq. (A3) in Ref.~\cite{Shao:2012fs}.

For parity-conserving decay process, all coefficients $A$'s equal to zero, and the distribution given in Eq.~\eqref{eq:jpsi} has 5 degrees of freedom (the $\lambda$'s). Since $SO(3)$ rotation has 3 degrees of freedom, there are two $SO(3)$ rotation-invariant observables, which can be chosen to be $W_2$ and $W_3$ in Eq.~\eqref{eq:jpsiRI3}, which become
\begin{subequations}
\begin{align}
\begin{split}
W_2 \to
\frac{1}{5\pi}
\frac{\lambda_\theta^2+3( \lambda_\phi^2 + \lambda_{\theta\phi}^2+ \lambda_{\perp \phi}^{2} + \lambda_{\perp \theta\phi}^{2})}{(3+\lambda_\theta)^2},
\end{split}\\
\begin{split}
W_3 \to
\frac{(\lambda_\theta+3\lambda_\phi)(2\lambda_\theta^2-6\lambda_\theta \lambda_\phi+9\lambda_{\theta \phi}^2)+9(\lambda_\theta \lambda_{\perp\theta\phi}^2 -2 \lambda_\theta \lambda_{\perp\phi}^2 +6\lambda_{\theta\phi}\lambda_{\perp\theta\phi} \lambda_{\perp\phi} -3 \lambda_\phi \lambda_{\perp\theta\phi}^{2})}{70\pi^2(3+\lambda_\theta)^3}.
\end{split}
\end{align}
\end{subequations}
The two $SO(2)$ rotation-invariant observables are still given in Eq.~\eqref{eq:jpsiRI2}.

For parity-conserving decay process, if there are frames in which both $\lambda_{\perp \phi}$ and $\lambda_{\perp \theta\phi}$ vanish\footnote{An example is one particle inclusive production with $x-z$ plane coinciding with production plane.}, $W_2$ and $W_3$ are further simplified in these frames to be
\begin{subequations}\label{eq:jpsiRIred}
\begin{align}
\begin{split}
W_2\rightarrow \frac{\lambda_\theta^2+3\lambda_\phi^2 + 3\lambda_{\theta\phi}^2}{(3+\lambda_\theta)^2},
\end{split}\\
\begin{split}
W_3\rightarrow
\frac{(\lambda_\theta+3\lambda_\phi)(2\lambda_\theta^2-6\lambda_\theta \lambda_\phi+9\lambda_{\theta \phi}^2)}{(3+\lambda_\theta)^3}.
\end{split}
\end{align}
\end{subequations}
It is easy to find that requiring the two quantites in Eq.~\eqref{eq:jpsiRIred} to be invariant is equivalent to requiring the two quantities in Eq.~\eqref{eq:jpsiRI2} to be invariant. Therefore, we conclude that these frames must be related by a rotation in $x-z$ plane.

\subsection{Spin-$2$ particle decay}

Since there are also higher spin bound states, such as $\chi_{c2}$ that can decay to a photon and a $J/\psi$, we give a brief discussion for spin-2 particle decay. We use the parameterization of the angular distribution given in Ref.~\cite{Shao:2012fs},
\begin{align}
\begin{split}
\frac{1}{\Gamma}\frac{d\Gamma}{d \Omega}=&\frac{1}{4\pi(1+\frac{\lambda_\theta}{3}+\frac{\lambda_{2\theta}}{5})} \Big[ 1+\lambda_{\theta}\cos^2\theta+\lambda_{2\theta}\cos^4\theta
+\lambda_{\theta\phi}\sin2\theta\cos\phi\\
&+\lambda_{2\theta\phi}\sin2\theta\sin^2\theta\cos\phi+\lambda_{\perp\theta\phi}\sin2\theta\sin\phi+
\lambda_{\perp2\theta\phi}\sin2\theta\sin^2\theta\sin\phi
\\
&+\lambda_{\phi}\sin^2\theta\cos2\phi+
\lambda_{2\phi}\sin^4\theta\cos2\phi+\lambda_{\perp\phi}\sin^2\theta\sin2\phi+\lambda_{\perp2\phi}\sin^4\theta\sin2\phi\\
&+
\lambda_{3\theta\phi}\sin2\theta\sin^2\theta\cos3\phi+\lambda_{\perp3\theta\phi}\sin2\theta\sin^2\theta\sin3\phi\\
&+\lambda_{4\phi}\sin^4\theta\cos4\phi+\lambda_{\perp4\phi}\sin^4\theta\sin4\phi\Big],\label{eq:spin2}
\end{split}
\end{align}
where parity conservation is assumed. With Eq.~\eqref{eq:RISO3}, we can construct many $SO(3)$ rotation invariants. Instead of listing all of them, we give explicit expression of one invariant that can be calculated from Eq.~\eqref{eq:RISO32},
\begin{align}\label{eq:spin2RI3}
\begin{split}
U_4=\frac{4\lambda_{2\theta}^2+10\lambda_{2\theta\phi}^2+5\lambda_{2\phi}^2+70\lambda_{3\theta\phi}^2+140\lambda_{4\phi}^2
+10\lambda_{\perp2\theta\phi}^2+5\lambda_{\perp2\phi}^2+70\lambda_{\perp3\theta\phi}^2+140\lambda_{\perp4\phi}^2
}{(15+5\lambda_\theta+3\lambda_{2\theta})^2}.
\end{split}
\end{align}
%
From Eq.~\eqref{eq:RISO2a}, we can also obtain the two $SO(2)$ rotation-invariant observables,
\begin{subequations}\label{eq:spin2RI2}
\begin{align}
\begin{split}
T_{2,0}=
-\frac{5}{14\sqrt{5\pi}}
\frac{7\lambda_\theta+6\lambda_{2\theta}+ 21\lambda_\phi + 18\lambda_{2\phi}}{15+5\lambda_\theta+3\lambda_{2\theta}},
\end{split}\\
\begin{split}
T_{4,0}=
\frac{1}{14\sqrt{\pi}}
\frac{3\lambda_{2\theta}-5\lambda_{2\phi} + 35\lambda_{4\phi}}{15+5\lambda_\theta+3\lambda_{2\theta}}.
\end{split}
\end{align}
\end{subequations}
It is straightforward to check that the complicated rotation-invariant observable $F_2$ defined in Eq. (B6) in Ref.~\cite{Shao:2012fs} can be obtained by linear combination of $T_{2,0}$ and $T_{4,0}$ defined in Eq.~\eqref{eq:spin2RI2}.

\section{Discussion}\label{sec:discussion}

The polarization of a particle with spin $J$ can be studied from the angular distribution of any of its decay products. In Eq.~\eqref{eq:YExpand}, we show that angular distribution can always be expanded by spherical harmonics $Y_{lm}(\theta,\phi)$ with $l\leq 2J$, with coefficients determined by spin density matrix of the decaying particle. These coefficients, however, also depend on the choice of a reference frame or a coordinate system, and a clear physical picture can be obtained from these coefficients only if a very clever reference frame/coordinate is chosen. In fact, an improper choice of the reference frame/coordinate could lead to artificial dependencies of the results on the kinematics and on the experimental acceptance \cite{Faccioli:2010kd}.  In contrast, rotation-invariant observables that we proposed in Eqs.~\eqref{eq:RISO32}, \eqref{eq:RISO3}, and \eqref{eq:RISO2a} do not depend on the selected reference frame/coordinate. Therefore they are expected to be better observables to extract the polarization information of the decaying particle and to test underlying theory.

A few $SO(2)$ rotation invariant observables have already been suggested in literature \cite{Faccioli:2010ej, Faccioli:2010ps, Palestini:2010xu, Shao:2012fs} and have been applied in the analysis of quarkonium polarization \cite{Chatrchyan:2012woa,Chatrchyan:2013cla}. We show that all these observables can be easily obtained with our method. In addition, our method provides more $SO(3)$ and $SO(2)$ rotation-invariant observables. For single inclusive production, all commonly-used frames are related by a $SO(2)$ rotation in the production plane. In this case the $SO(2)$ rotation-invariant observables are adequate for the comparison between prediction and measurements, and between different measurements. For a general process, such as associate productions, the most important frame choices may not be related by a $SO(2)$ rotation. Then the $SO(3)$ rotation-invariant observables are necessary.

In section~\ref{sec:application}, we also calculate the explicit expressions of all rotation-invariant observables for the decay process of spin-1/2 particle and spin-1 particles, and some rotation-invariant observables for the decay process of a spin-2 particle. These expressions can be used to check the possibility of unaddressed systematic uncertainty, caused by the detector acceptence or by the event selection criteria \cite{Faccioli:2010kd}.  They can facilitate the comparison between existing analyses of polarization. They are also a robust test of the correctness in the future polarization analysis.

\begin{acknowledgements}

H.Z. thank E. Braaten, J. Huang and B. Gui for helpful discussions. J.Q. and H.Z. were supported by the U.S. Department of Energy under the contract DE-AC05-06OR23177 and grant DE-SC0011726, respectively.

\end{acknowledgements}

\providecommand{\href}[2]{#2}\begingroup\raggedright\endgroup

\end{document}